\documentclass[preprint,showpacs,superscriptaddress,aps]{revtex4-1}
\usepackage{amssymb}
\usepackage{amsmath,amssymb,graphicx}
\usepackage{graphicx}% Include figure files
\usepackage{dcolumn}% Align table columns on decimal point
\usepackage{bm}% bold math
\def\be{\begin{equation}}
\def\ee{\end{equation}}
\def\bea{\begin{eqnarray}}
\def\eea{\end{eqnarray}}

\def\lsim{\raise0.3ex\hbox{$\;<$\kern-0.75em\raise-1.1ex\hbox{$\sim\;$}}}
\def\gsim{\raise0.3ex\hbox{$\;>$\kern-0.75em\raise-1.1ex\hbox{$\sim\;$}}}

\usepackage{pstricks}
\begin{document}

\title{$CP$ violation in
$D^0 \to K^+ \pi^- $ }

\author{David Delepine}
\email{delepine@fisica.ugto.mx}
\affiliation{{\fontsize{10}{10}\selectfont{Division de Ciencias
e Ingenier\'ias,  Universidad de Guanajuato, C.P. 37150, Le\'on, Guanajuato, M\'exico.}}}

\author{Gaber Faisel}
\email{gaberfaisel@sdu.edu.tr}

\affiliation{{\fontsize{10}{10}\selectfont{Department of Physics,
Faculty of Arts and Sciences, S\"uleyman Demirel University,
Isparta, Turkey 32260.}}}

\author{ Carlos A. Ramirez}
\email{jpjdramirez@yahoo.com}
\affiliation{{\fontsize{10}{10}\selectfont{Depto. de F\'isica,
Universidad de los Andes, A. A. 4976-12340, Bogot\'a, Colombia.}}}

\begin{center}

\begin{abstract}
 In this paper we study the direct CP asymmetry of
the doubly Cabibbo-suppressed decay mode $D^0 \to K^+ \pi^- $
within standard model and two Higgs doublet model with generic
Yukawa structure. In the standard model we derive the corrections
to the tree level amplitude, generated from the box and di-penguin
diagrams, required for generating the weak CP violating phases. We
show that these phases are so tiny leading to a direct CP
asymmetry of order $10^{-9}$. Regarding the two Higgs doublet
model with generic Yukawa structure we derive the Wilson
coefficients relevant to $D^0 \to K^+ \pi^- $.  After taking into
account all constraints on the parameter space of the model we
show that charged Higgs couplings to quarks can lead to a direct
CP asymmetry of order $10^{-3}$ which is $6$ orders of magnitude
larger than the standard model prediction.

\end{abstract}
\end{center}

\maketitle
\section{Introduction}

Up to now,  no signals for new particles  beyond the standard model (SM) have been seen in
colliders.  New Physics (NP) can be also probed through indirect searches in
colliders, for instances searching for signals of flavor
violation in the quark sector forbidden in the SM, lepton number violation and CP
violation beyond the one predicted by SM.

In the SM the origin of CP violation is the
Cabibbo-Kobayashi-Maskawa (CKM) matrix describing the quark mixing
\cite{Cabibbo:1963yz,Kobayashi:1973fv}. The presence of some
complex elements in the CKM matrix allows  CP violation that has
been observed in kaon and B mesons
\cite{Christenson:1964fg,Aubert:2004qm,Aaij:2013iua,Aaij:2012kz}.
Moreover, recently LHCb has reported the first measurement of CP
violation in the baryon sector using the baryonic decay mode
$\Lambda_b \to p \, \pi^- \pi^+ \pi^-$
decays\cite{Raaij,Dordei:2017cjf}. On the other hand, in the D
sector, great experimental progress has been achieved in the last
decade. The $D^0 -\bar D^0$ mixing was discovered in 2007 after
combining the results from BABAR \cite{Aubert:2007wf}, Belle
\cite{Staric:2007dt} and CDF \cite{Aaltonen:2007ac}. Later, at
LHCb, the mixing has been firmly established after the first
experimental observations of the slow mixing rate of the $D^0
-\bar D^0$ oscillations \cite{Aaij:2012nva}. Despite this progress there is
still no experimental evidence for direct CP violation in charm.
The first two full years data taking at LHCb are consistent with
CP conservation in charm \cite{Smith:2016dsz}.

CP violating processes involving the up-type quark are expected to
be seen only in the charmed mesons. In this sector CP violation
within SM is expected to be small because the relevant combination
of elements of the CKM matrix is of order $10^{-3}$
\cite{Nierste:2017cgo}. Generally two body non-leptonic D decays
can be classified into  Cabibbo-Favored (CF), single
Cabibbo-suppressed (SCS) and Double Cabibbo Suppressed (DCS). This
classification is based on the power of the suppression factor
$\lambda \simeq |V_{us}| \simeq |V_{cd}|$ which appears in their
amplitudes \cite{Nierste:2017cgo}.

 Within SM CP-asymmetry of order $10^{-3}$ has been predicted in
some  SCS decay modes \cite{Bhattacharya:2012ah}. More SCS decay
modes have been investigated in the framework of the SM.  The
results showed that a larger CP-asymmetry of order $10^{-2}$ can
be obtained for the decay modes $D^0 \to K_s K_s$
\cite{Nierste:2015zra}. In a recent study  of the SCS decays $D^0
\to K_s K^{*^0}$ and $D^0 \to K_s \bar K^{*^0}$ the direct CP
asymmetry have been estimated in the SM to be as large as $ 3
\times 10^{-3}$ \cite{Nierste:2017cua}. We turn now to CF two body
non-leptonic D decays. In a previous study we showed that the
direct CP asymmetry of $D^0 \to K^- \pi^+ $  can be less than or
equal $ 1.4 \times 10^{-10}$ in the framework of the SM
\cite{Delepine:2012xw}. This almost vanishing asymmetry favors
this decay mode to be smoking gun for NP beyond SM. In fact one
expects to have similar situation for the DCS decay mode $D^0 \to
K^+ \pi^- $. So the objectives of this study is to give a
prediction of the direct CP asymmetry of this process in SM and to
explore NP contribution to this asymmetry arising from two Higgs
doublet model with generic Yukawa structure.

Simple extensions of the SM include the two Higgs doublet models
(2HDM)\cite{Haber:1978jt,Abbott:1979dt}. These models keep the
gauge  structure of the SM untouched. They only extend the scalar
sector by adding new scalars. 2HDM can be classified to several
types according to their couplings to quarks and leptons. For
instances, 2HDM type I, II or III (for a review see ref.
\cite{Branco:2011iw}). The 2HDM III  has complex couplings to
quarks. As a consequence these couplings are relevant for
generating the desired CP violating weak phases. Other motivation
for 2HDM III includes their ability to explain $B\to D\tau\nu$,
$B\to D^*\tau\nu$ and $B\to \tau\nu$ simultaneously while other
types such as 2HDM I and 2HDM II cannot \cite{Crivellin:2012ye}.

 This paper is organized as follows. In Sec.~\ref{sM},
we study the SM contribution to the amplitude of the decay mode
$D^0 \to K^+ \pi^-$. At tree-level the amplitude has no source of
the weak CP violating phases required for non-vanishing direct CP
asymmetry. Accordingly, we consider the loop-level and derive the
contributions generated from box and di-penguin diagrams. In
addition, we calculate the SM prediction of the direct CP
asymmetry. In Sec.~\ref{Hig} we derive the contributions relevant
to the Wilson coefficients of $D^0 \to K^+ \pi^-$ originated from
a charged Higgs couplings to the quarks in a two Higgs doublet
model with generic Yukawa structure. Finally, we give our
conclusion in Sec.~\ref{sec:conclusion}.

\section{ Direct CP asymmetry of $D^0 \to K^+ \pi^-$ within SM \label{sM}}

Within SM the weak effective Hamiltonian governing the decay
process  $D^0 \to K^+ \pi^-$ can be written as

\begin{eqnarray}
{\cal H}^{SM}_{\rm eff.} &=&
{G_F\over\sqrt{2}}V_{cd}^*V_{us}\left(c_1\bar d \gamma_\mu c_L\bar
u \gamma^\mu s_L
+c_2\bar u \gamma_\mu c_L\bar d\gamma^\mu s_L\right)+{\rm h.c.} \nonumber\\
&=& {G_F\over\sqrt{2}}V_{cd}^*V_{us}\left(a_1{\cal O}_1+a_2{\cal
O}_2\right)+{\rm h.c.} \label{SMH}
\end{eqnarray}
where $a_1\equiv c_1+ c_2/N_c $ and $a_2\equiv c_2-c_1/N_C$ and
$N_C$ is the color number.  In naive factorization approximation
(NFA) the amplitude of  $D^0\to K^+\pi^-$ can be written as
\begin{eqnarray}
A_{D^0\to K^+\pi^-} &=&-i{G_F\over \sqrt{2}}V_{cd}^*V_{us}
\left[a_1X^{K^+}_{D^0 \pi^-} +a_2X^{D^0}_{K^+\pi^-} \right],
\label{am1}
\end{eqnarray}

\begin{figure}[tbhp]
  % Requires \usepackage{graphicx}
  \includegraphics[width=7.5cm]{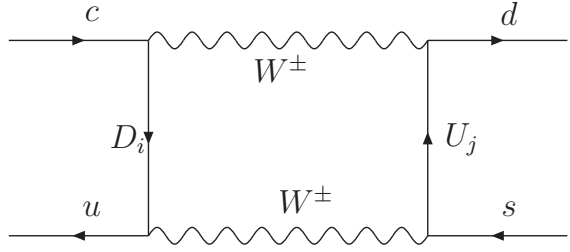}\hspace{1.cm}

  \caption{Feynman diagram for DCS processes: Box  contribution.}
\end{figure}

 where $ X^{P_1}_{P_2P_3}$ is given by
\begin{eqnarray}
X^{P_1}_{P_2P_3}= if_{P_1}\Delta_{P_2P_3}^2
F_0^{P_2P_3}(m_{P_1}^2),\  \Delta_{P_2P_3}^2=m_{P_2}^2-m_{P_3}^2
\end{eqnarray}
here $f_{P}$ is the $P$ meson  decay constant and $F_0^{P_2 P_3}$
is the form factor.

In NFA, there is no source for the strong CP conserving phases
required for having non vanishing direct CP aymmetries.
Consequently this factorization approximation is irrelevant to the
study of CP violation. On the other hand the mass of the charm
quark is not heavy enough to allow for a sensible heavy quark
expansion, such as in QCD factorization and soft collinear
effective theory, and it is not light enough for the application
of chiral perturbation theory \cite{Cheng:2010ry}.  A possible
approach to study charm decays in a model-independent way is the
so called the diagrammatic approach \cite{Chau:1982da,
Chau:1986du,Chau:1987tk,Chau:1989tk,Buccella:1994nf,
Cheng:2010ry}. Within this approach, the amplitude is decomposed
into  parts corresponding to generic quark diagrams according to
the topologies of weak interactions. For each one of these
topological diagrams, the related magnitude and relative strong
phase can be extracted from the data without making further
assumptions, apart from flavor SU(3) symmetry \cite{Cheng:2010ry}.

 In the diagrammatic approach the amplitude of the decay
process  $D^0 \to K^+ \pi^-$ can be written as \cite{Cheng:2010ry}
\be {\mathcal A}_{D^0\to K^+\pi^-}= V^*_{cd}V_{us}(T''+E'')
\label{HigsTt1}\ee where $T''$ is the tree level color-allowed
external W-emission quark diagram and $E''$ is the W-exchange
quark diagram. Their magnitudes and their strong phases can be
found in Ref.\cite{Cheng:2010ry}. Comparing  Eqs.(\ref{am1}) and
(\ref{HigsTt1})  we get \bea T''= \frac{G_F}{\sqrt{2}} a_1
f_{K}(m^2_D-m^2_\pi)F^{D\pi}_0(m^2_{K})\nonumber\\
E''= \frac{G_F}{\sqrt{2}} a_2
f_D(m^2_K-m^2_{\pi})F^{K\pi}_0(m^2_D)\eea

\begin{figure}[tbhp]
  % Requires \usepackage{graphicx}
  \includegraphics[width=8.5cm]{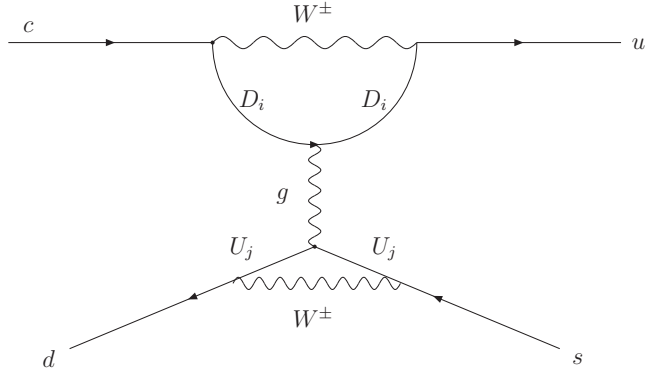}
 \caption{Feynman diagram for DCS processes:di-penguins contribution.\label{fi2}}
\end{figure}

 In the SM the Wilson coefficients $a_1$ and $a_2$ and
 the CKM elements $V_{cd}$ and  $V_{us}$ are all real. Thus at
 this level there is no source for the CP violating weak phases
 required for non vanishing CP asymmetry. Possible CP violating weak
 phases can be generated through box and di-penguin diagrams in
 fig.1 and fig.2. The box contribution to the total SM weak effective Hamiltonian
 can be expressed as

 \begin{eqnarray}
\Delta{\cal H}^{box}  &=& \frac{G_F^2m_W^2}{ 2\pi^2}
V_{cD}^*V_{uD} V_{Ud}^*V_{Us}f(x_U,\ x_D)
\bar u\gamma_\mu c_L\bar d \gamma^\mu s_L \nonumber\\
&=&\frac{G_F^2m_W^2}{ 2\pi^2} \lambda^D_{cu}\lambda^U_{d s}f(x_U,\ x_D){\cal O}_2 \nonumber\\
&=&\frac{G_F^2m_W^2}{ 2\pi^2} {\cal B}_x {\cal O}_2
\end{eqnarray}
where
\begin{eqnarray}
{\cal B}_x &=& \lambda^D_{cu}\lambda^U_{d s}f(x_U,\ x_D) \\
 &=&  V_{cd}^*V_{ud}\left( V_{ud}^*V_{u s}f_{ud}+V_{cd}^*V_{cs}f_{cd}+V_{td}^*V_{ts}f_{td}  \right)
+ V_{cs}^*V_{us}\left( V_{ud}^*V_{u s}f_{us}+V_{c d}^*V_{c
s}f_{cs}+V_{td}^*V_{t s}f_{ts} \right) \nonumber
\\&+& V_{cb}^*V_{ub}\left( V_{u d}^*V_{u s}f_{ub}+V_{cd}^*V_{c
s}f_{cb}+V_{t d}^*V_{t s}f_{tb} \right) \end{eqnarray} where
$U=u,\ c,\ t$, $D=d,\ s,\ b$ and $\lambda_{DD'}^U \equiv
V_{UD}^*V_{UD'}$, $\lambda_{UU'}^D \equiv V_{UD}^*V_{U'D}$,
$x_q=(m_q/m_W)^2$ and $f_{UD} \equiv f(x_U,x_D)$ \cite{inami}
\begin{eqnarray}
f(x,\ y) ={7xy-4\over 4(1-x)(1-y)} +{1\over x-y}\left[ {y^2\log
y\over (1-y)^2}\left(1-2x+{xy\over 4}\right)-  {x^2\log x\over
(1-x)^2}\left(1-2y+{xy\over 4}\right)   \right] \nonumber
\end{eqnarray}

Clearly ${\cal B}_x$ will be a complex number due to the presence
of the complex CKM elements. Thus the desired CP violating weak
phases are generated through the box contribution to the weak
effective Hamiltonian.

The other contribution to the total SM weak effective Hamiltonian
is the di-penguin contribution generated via diagram in
Fig.(\ref{fi2}) and can be written as

\begin{eqnarray}
\Delta {\cal H}^{dipeng.} &=& -{G_F^2\alpha_S\over 8\pi^3}\left[
\lambda^D_{cu}  E_0(x_D)\right] \left[ \lambda^U_{d s}
E_0(x_U)\right]    \bar d \gamma_\mu T^a s_L
\left(g^{\mu \nu} \Box-\partial^\mu\partial^\nu \right) \bar u \gamma_\nu T^a c_L\nonumber \\
&=&  -{G_F^2\alpha_S\over 8\pi^3} {\cal P}_g \, \bar d \gamma_\mu
T^a s_L \left(g^{\mu \nu} \Box-\partial^\mu\partial^\nu \right)
\bar u
\gamma_\nu T^a c_L \nonumber \\
&\equiv & {G_F^2\alpha_S\over  16\pi^3} {\cal P}_g \,{\cal
O}\end{eqnarray}

where $T^a$ are the generator of $SU(3)_C$ and the Inami function
is given by
\begin{eqnarray}
E_0(x) &=&     {1\over 12(1-x)^4}\left[
x(1-x)(18-11x-x^2)-2(4-16x+9x^2)\log(x)\right]
\end{eqnarray}

The quantity ${\cal P}_g$ can be expressed in terms of CKM
elements and Inami function as
\begin{eqnarray}
{\cal P}_g&\equiv & \left[  \lambda^D_{cu}  E_0(x_D)\right] \left[
\lambda^U_{ds} E_0(x_U)\right]=\left[
V_{cs}^*V_{us}\left(E_0(x_s)-E_0(x_d)\right)+ V_{cb}^*
V_{ub}\left( E_0(x_b)-E_0(x_d) \right)   \right]
\nonumber \\
&& \left[ V_{cs}V_{cd}^*  \left(E_0(x_c)-E_0(x_u)\right) +
V_{ts}V_{td}^* \left(E_0(x_t)-E_0(x_u)\right)\right]
\label{pge}\end{eqnarray}

As can be seen from Eq.(\ref{pge}) ${\cal P}_g$ is a complex
number due to the presence of the complex CKM elements. Thus
di-penguin contribution generates CP violating weak phases. We now
proceed to reduce the operator ${\cal O}$ and find that
\begin{eqnarray}
{\cal O} &=& \bar d \gamma_\mu T^a s_L \left(g^{\mu \nu}
\Box-\partial^\mu\partial^\nu \right) \bar u \gamma_\nu T^a c_L  =
\bar d \gamma_\mu T^a s_L \Box \left(\bar u \gamma^\nu T^a
c_L\right) + \bar d \partial \hskip-0.2cm /\ T^a s_L  \bar u
\partial \hskip-0.2cm /\ T^a c_L
\nonumber \\
&=& -q^2 \bar d \gamma_\mu T^a s_L \bar u \gamma^\mu T^a c_L-
\left(m_d \bar d T^a s_{S-P}+ m_s \bar d T^a s_{S+P} \right)
\cdot \left(m_c \bar u T^a c_{S+P}+m_u \bar u T^a c_{S-P}\right) \nonumber \\
 &&-q^2 \bar d \gamma_\mu T^a s_L \bar u \gamma^\mu T^a c_L -m_d m_c \bar d T^a s_L
 \bar u T^a c_R - m_s m_u\bar d T^a s_R \bar u T^a c_L \nonumber \\
 && -m_d m_u\bar d  T^a s_L \bar u T^a c_L-m_s m_c\bar d T^a s_R \bar u T^a c_R
\end{eqnarray}

This expression can be further simplified using
\begin{eqnarray}
\bar d \gamma_\mu T^a s_L \bar u \gamma^\mu T^a c_L &=& {1\over 2}\left({\cal O}_1-{1\over N_C}{\cal O}_2\right) \nonumber \\
\bar d T^a s_L \bar u T^a c_R &=&-{1\over 4}\bar d \gamma_\mu c_R \bar u \gamma^\mu s_L -{1\over 2 N_C}\bar d s_L \bar u c_R   \nonumber \\
\bar d T^a s_R \bar u T^a c_L &=& -{1\over 4}\bar d \gamma_\mu c_L \bar u \gamma^\mu s_R -{1\over 2 N_C}\bar d s_R \bar u c_L \nonumber \\
\bar d T^a s_L \bar u T^a c_L &=&-{1\over 4}\bar d c_L \bar u s_L
-{1\over 16}\bar d\sigma_{\mu\nu} c_L \bar u \sigma^{\mu\nu} s_L
-{1\over 2 N_C}\bar d s_L \bar u c_L \nonumber \\
\bar d T^a s_R \bar u T^a c_R &=&-{1\over 4}\bar d c_R \bar u s_R
-{1\over 16}\bar d\sigma_{\mu\nu} c_R \bar u \sigma^{\mu\nu}
s_R-{1\over 2 N_C}\bar d s_R \bar u c_R
\end{eqnarray}

Upon taking the expectation values, we obtain
\begin{eqnarray}
\left<{\cal O}\right> &=&-q^2 \left<\bar d \gamma_\mu T^a s_L \bar
u \gamma^\mu T^a c_L\right> -m_d m_c \left<\bar d T^a s_L \bar u
T^a c_R\right>
-m_s m_u\left<\bar d T^a s_R \bar u T^a c_L\right> \nonumber \\
&&-m_d m_u\left<\bar d T^a s_L \bar u T^a c_L\right>-m_s
m_c\left<\bar d T^a s_R \bar u T^a c_R\right>
\nonumber \\
&\simeq &-{q^2\over 2}\left(1-{1\over
N^2}\right)X^{K^+}_{D^0\pi^-}+{m_d m_c\over 4}\left(1-{1\over
N}\right)X^{K^+}_{D^0\pi^-}+{5m_s \over 8N m_d}m_D^2X^{D^0}_{K^+
\pi^-}
\end{eqnarray}
where  $q^2$ is the gluon momentum. For the decay $D^0\to
K^+\pi^-$, one can approximate $q^2=(p_c \mp p_u)^2=(p_s\pm
p_d)^2\simeq (p_D-p_\pi/2)^2=(m_D^2+m_K^2)/2+3m_\pi^2/4$, by
assuming that $p_c\simeq p_D$ and $p_u\simeq p_\pi/2$. Finally,
including both box and di-penguin contributions leads to a
modification of the Wilson coefficients $a_1$ and $a_2$ as $a_i\to
a_i+\Delta a_i$ for $i=1,2$ and $\Delta a_i$ are given by

\begin{eqnarray}
\Delta a_1 &=& -{G_Fm_W^2\over \sqrt{2}\ \pi^2V_{cd}^*V_{us}N
}\,{\cal B}_x - {G_F \alpha_S\over 4\sqrt{2}
\pi^3V_{cd}V_{ud}^*}\left[{q^2\over 2}\left(1-{1\over N^2}\right)
-{m_c m_d\over 4}\left(1-{1\over N}\right)  \right] {\cal P}_g \nonumber \\
\Delta a_2 &=& -{G_Fm_W^2\over \sqrt{2}\ \pi^2V_{cd}^*V_{us} }
\,{\cal B}_x  - {G_F \alpha_S\over 4\sqrt{2} \pi^3V_{c
d}V_{ud}^*}{5m_s m_D^2\over 8N m_d} {\cal P}_g \label{dela}
\end{eqnarray}

For the numerical analysis we need to specify the values of the
CKM elements. The CKM matrix can be fully determined by the
measurement of four independent parameters. A convenient
determination is from tree-level charged current decays only,
which can be used to find the mixing angles\cite{Blanke:2016uwb}.
These are $|V_{us}|$, $|V_{cb}|$ and $|V_{ub}|$ in addition to the
CP violating angle $\gamma$ of the unitarity triangle. For the
$CKM$ matrix element $V_{td}$ we follow Ref.\cite{Buras:2012jb}
and evaluate it using

\be V_{td} = |V_{us}| |V_{cb} | R_t\, e^{- i \beta},\ee

where

\be R_t = \sqrt{1 + R_b^2 - 2 R_b \cos\gamma},\,\, R_b= (1 -
\frac{\lambda^2}{2}) \frac{1}{\lambda}
\frac{|V_{ub}|}{|V_{cb}|},\,\,  \cot\beta= \frac{1 - R_b
\cos\gamma}{R_b \sin\gamma}\ee

 From kaon decays we have \cite{Olive:2016xmw}
$|V_{us}|=0.2248\pm 0.0006$. The accuracy of the current
experimental determination of $\gamma$ by the LHCb collaboration
\cite{Aaij:2016kjh} $\gamma=(72.2^{\,+6.8}_{\,-7.2})^\circ$. The
situation for $|V_{cb}|$ and $|V_{ub}|$ is quite unsatisfactory.
This is due to the significant discrepancies in the determined
values using inclusive or exclusive decays. The uncertainties in
extracting $|V_{ub}|$ from inclusive and exclusive decays are
different to a large extent. Exclusively, the most precise
determination of $|V_{ub}|$ ($|V_{cb}|$) is obtained from the
decay $B\to \pi\ell \nu$ ($B\to D^*\ell \nu$)
\cite{Blanke:2016uwb,Lattice:2015tia}
(\cite{Blanke:2016uwb,Bailey:2014tva})

\be |V^{\,excl.}_{ub}|=(3.72\pm 0.16)\times
10^{-3},\,\,\,\,\,\,\,\,\,\,\, |V^{\,excl.}_{cb}|=(39.04\pm
0.75)\times 10^{-3}\ee

On the other hand we have from the inclusive decay $B\to X_u\ell
\nu$ ($B\to X_c\ell \nu$) \cite{Olive:2016xmw}
(\cite{Gambino:2016jkc})

\be |V^{\,incl.}_{ub}|=(4.41\pm 0.15^{+0.15}_{-0.19})\times
10^{-3},\,\,\,\,\,\,\,\,|V^{\,incl.}_{cb}|=(42.00\pm
0.65^{+0.15}_{-0.19})\times 10^{-3} \ee

A combination of the inclusive and exclusive of $|V_{ub}|$ and
$|V_{cb}|$ determinations is quoted \cite{Kowalewski:2008zz}

\be |V_{ub}|=(4.09\pm 0.39)\times
10^{-3},\,\,\,\,\,|V_{cb}|=(40.5\pm 1.5)\times 10^{-3}\ee

Having these input values we can now obtain the numerical values
of $\Delta a_{1}$ and $\Delta a_{2}$ given in Eq.(\ref{dela}). We
list their predictions in Table\ref{SMpr}. Clearly from
Table\ref{SMpr} the predicted values of $\Delta a_{1}$ and $\Delta
a_{2}$ corresponding to the inclusive, the exclusive and the
combined input values of the CKM elements $|V_{cb}|$ and
$|V_{ub}|$ are approximately equal. Thus in the following we give
our predictions corresponding to the combined input values of
$|V_{cb}|$ and $|V_{ub}|$.

\begin{table}
\begin{center}
\begin{tabular}{|c|c|c|c|}
  \hline
  % after \\: \hline or \cline{col1-col2} \cline{col3-col4} ...
   & inclusive & exclusive & combined \\
  \hline
  $\Delta a_1$ & $2.2 \cdot 10^{-7} {\rm e}^{179.68^\circ\, i}$ & $ 2.2 \cdot 10^{-7} {\rm
e}^{179.74^\circ\, i}$  & $2.2 \cdot 10^{-7} {\rm
e}^{179.71^\circ\, i}$  \\
$\Delta a_2$ & $ 2.2 \cdot 10^{-6} {\rm
e}^{179.88^\circ\, i} $ & $2.2 \cdot 10^{-6} {\rm e}^{179.90^\circ\, i}$ & $2.2 \cdot 10^{-6} {\rm e}^{179.89^\circ\, i}$ \\
\hline
\end{tabular}
\end{center}
\caption{ Predictions for $ \Delta a_1$ and $\Delta a_2$
corresponding to the inclusive, the exclusive and the combined
input values of the CKM elements $|V_{cb}|$ and
$|V_{ub}|$.}\label{SMpr}
\end{table}

The direct CP asymmetry of $D^0 \to K^+ \pi^-$ can be expressed as

\begin{eqnarray}
A_{CP} &=& {|A|^2-|\bar A|^2 \over |A|^2+|\bar A|^2 }= {2 r
\sin(\phi_2-\phi_1)\sin(\alpha)\over |1+ r|^2 } = \kappa\,
\sin(\phi_2-\phi_1)\label{acp}
\end{eqnarray}

where we have defined \be \kappa = {2 r\sin(\alpha)\over |1+ r|^2
}\ee

with $r=|E''/T''|$ and $\alpha= \alpha_{E''}- \alpha_{T''}$.  The
phases $\alpha_{E''}$ and $\alpha_{T''}$ are the strong phase of
the amplitudes $E''$ and $T''$ respectively. The weak phases
$\phi_1$ and $\phi_2$ are defined through \be \phi_i =
\tan^{-1}\big( \frac{|\Delta a_i|\sin\Delta \phi_i}{a_i+|\Delta
a_i|\cos\Delta \phi_i}\big) \ee where $\Delta \phi_i$ is the phase
of $\Delta a_i$. With $a_1= 1.2 \pm 0.1$ and $a_2 = - 0.5 \pm 0.1$
and $ \Delta a_1$ and $\Delta a_2$ given in the last coulomb of
Table \ref{SMpr} we find that $\sin(\phi_2-\phi_1) \simeq -9\times
10^{-9}$ and hence

\begin{eqnarray}
A_{CP}\simeq -9\times 10^{-9} \kappa
\end{eqnarray}

For $T'' = (3.14 \pm 0.06)\times 10^{-6} $ and $E'' =
(1.53^{+0.07}_{-0.08})\times 10^{-6} e^{i(122\pm 2)\circ}$
\cite{Cheng:2010ry} we find that $\kappa \simeq 0.37$ and thus $
A_{CP} \simeq - 3.36 \times 10^{-9}$. Clearly the predicted direct
CP asymmetry within SM is so tiny.

\section{Models with Charged Higgs contributions}\label{Hig}
In  2HDM III the physical mass eigenstates are $H_0$ (heavy
CP-even Higgs), $h_0$ (light CP-even Higgs) and $A_0$ (CP-odd
Higgs) and $H^{\pm}$. In this model both Higgs doublets can couple
to up-type and down-type  quarks. As a consequence the couplings
of the neutral Higgs mass eigenstates can induce flavor violation
in Neutral Currents at tree-level. In the down sector these flavor
violating couplings are stringently constrained from flavor
changing neutral current processes
\cite{Crivellin:2012ye,Crivellin:2013wna}. Thus in the following
we consider only charged Higgs couplings to quarks that can be
expressed as \cite{Crivellin:2010er,Crivellin:2012ye}:
\begin{equation}
\mathcal{L}^{eff}_{H^\pm} = \bar{u}_f {\Gamma_{u_f d_i
}^{H^\pm\,LR\,\rm{eff} } }P_R d_i
+ \bar{u}_f {\Gamma_{u_f d_i }^{H^\pm\,RL\,\rm{eff} } }P_L d_i\, ,\\
 \label{Higgs-vertex}
\end{equation}
where \bea {\Gamma_{u_f d_i }^{H^\pm\,LR\,\rm{eff} } } &=&
\sum\limits_{j = 1}^3 {\sin\beta\, V_{fj} \left( \frac{m_{d_i
}}{v_d} \delta_{ji}-
  \epsilon^{ d}_{ji}\tan\beta \right), }
\nonumber\\
{\Gamma_{u_f d_i }^{H^ \pm\,RL\,\rm{eff} } } &=& \sum\limits_{j =
1}^3 {\cos\beta\,  \left( \frac{m_{u_f }}{v_u} \delta_{jf}-
  \epsilon^{ u\star}_{jf}\tan\beta \right)V_{ji}}
 \label{Higgsv}
\eea Here $v_u$ and $v_d$ denote the vacuum expectations values of
the neutral component of the  Higgs doublets,  tan $\beta =
v_u/v_d$ and $V$ is the CKM matrix. Applying the Feynman-rules
given in Eq.(\ref{Higgs-vertex}) allows us to derive the effective
Hamiltonian, resulting from the tree level exchanging charged
Higgs diagram, that governs the process under consideration. The
effective Hamiltonian can be expressed as \be {\mathcal
H}^{H^\pm}_{eff}= \frac{ G_F}{\sqrt{2}}V^*_{cd}V_{us} \sum^4_{i=1}
C^H_i(\mu) Q^H_i(\mu),\ee

The Wilson coefficients $C^H_i$ are obtained by perturbative QCD
running from $M_{H^{\pm}}$ scale to the scale $\mu\simeq m_c$
relevant for hadronic decay and $Q^H_i$ are the relevant local
operators at  $\mu\simeq m_c$. The operators are given as %
\bea
Q^H_1 &=&(\bar{d} P_R c)(\bar{u} P_L s),\nonumber\\
Q^H_2 &=&(\bar{d} P_L c)(\bar{u} P_R s),\nonumber\\
Q^H_3 &=&(\bar{d} P_L c)(\bar{u} P_L s),\nonumber\\
Q^H_4 &=&(\bar{d} P_R c)(\bar{u} P_R s),
 \eea
 Their corresponding Wilson coefficients $C^H_i$, at $ \mu= m_H$ scale,
can be expressed as

\begin{eqnarray}
C^H_1 &=& \frac {\sqrt{2}  \cos^2\beta }{ G_F V^*_{cd}V_{us}
 m^2_H} \bigg(\frac{m_u V_{us} }{v_u}  - \sum\limits_{j = 1}^3
V_{j2} \epsilon^{ u\star}_{j1}\tan\beta \bigg)\bigg( \frac{m_c
V^*_{cd}}{v_u} - \sum\limits_{k= 1}^3 {V^{\star}_{k1}}\epsilon^{
u}_{k2}\tan\beta
\bigg),\nonumber\\
C^H_2 &=& \frac {\sqrt{2}  \sin^2\beta }{ G_F V^*_{cd}V_{us}
 m^2_H} \bigg( \frac{m_s  V_{us}}{v_d} - \sum\limits_{j = 1}^3
V_{1j} \epsilon^{ d}_{j2}\tan\beta \bigg)\bigg( \frac{m_d
V^*_{cd}}{v_d} -\sum\limits_{k= 1}^3 V^{\star}_{2k}\epsilon^{
d\star}_{k1}\tan\beta
\bigg),\nonumber\\
 C^H_3 &=& \frac {\sin 2\beta}{ \sqrt{2}\, G_F V^*_{cd}V_{us}
 m^2_H}\bigg(\frac{m_u  V_{us} }{v_u} - \sum\limits_{j = 1}^3
V_{j2} \epsilon^{ u\star}_{j1}\tan\beta \bigg)\bigg( \frac{m_d
V^*_{cd}}{v_d} -\sum\limits_{k= 1}^3 V^{\star}_{2k}\epsilon^{
d\star}_{k1}\tan\beta
\bigg),\nonumber\\
C^H_4 &=& \frac {\sin2\beta}{\sqrt{2}\, G_F V^*_{cd}V_{us}
 m^2_H}\bigg( \frac{m_s V_{us}}{v_d}  - \sum\limits_{j = 1}^3
V_{1j} \epsilon^{ d}_{j2}\tan\beta \bigg)\bigg( \frac{m_c
V^*_{cd}}{v_u}- \sum\limits_{k= 1}^3 {V^{\star}_{k1}}\epsilon^{
u}_{k2}\tan\beta \bigg),
 \label{Higgsw}
\end{eqnarray}

Having  deriving  the effective Hamiltonian we proceed now to
discuss the experimental constraints on the flavor-changing
parameters $\epsilon^{ q}_{ij}$, for $q=d,u$, appear in the Wilson
coefficients. We start first by discussing the constraints on
$\epsilon^{d}_{ij}$. For $i\neq j$ case we find that
$\epsilon^d_{ij}$ are stringently  constrained from FCNC processes
because of the tree-level neutral Higgs exchange
\cite{Crivellin:2012ye,Crivellin:2013wna}. Thus, we are left with
only $\epsilon^d_{11},\epsilon^d_{22}$.  These couplings can be
constrained upon applying the naturalness criterion of 't Hooft to
the quark masses \cite{Crivellin:2012ye}. Based on this criterion,
the smallness of a quantity is only natural if a symmetry is
gained in the limit in which this quantity is zero
\cite{Crivellin:2012ye}. As a result it is unnatural to have large
accidental cancellations without a symmetry forcing these
cancellations. Applying this criterion to the quark masses in the
2HDM III we get \cite{Crivellin:2012ye}

\begin{eqnarray}
|\epsilon^{d(u)}_{ij}|\leq \frac{\left|V_{ij}\right|\,{\rm max
}\left[m_{d_i(u_i)},m_{d_j(u_j)}\right]}{|v_{u(d)}|}\,.\label{constr}
\end{eqnarray}

This bound shows that $\epsilon^d_{11},\epsilon^d_{22}$  will be
severely constrained by their small quark masses. This is also the
case for the coupling $\epsilon^u_{11}$. Thus we conclude that all
the couplings $\epsilon^{d}_{ij}$ and $\epsilon^u_{11}$ that
appear in the Wilson coefficients in Eq.(\ref{Higgsw}) will lead
to negligible effects and hence can be safely dropped. Thus, to a
good approximation, we can write

\begin{eqnarray}
C^H_1 &\simeq& \frac {\sqrt{2}  \cos^2\beta }{ G_F V^*_{cd}V_{us}
 m^2_H} \bigg( \sum\limits_{j = 2}^3
V_{j2} \epsilon^{ u\star}_{j1}\tan\beta \bigg)\bigg(
\sum\limits_{k= 1}^2 {V^{\star}_{k1}}\epsilon^{ u}_{k2}\tan\beta
\bigg),\nonumber\\
C^H_4 &\simeq& - \frac {\sin2\beta m_s}{\sqrt{2}\, G_F V^*_{cd}
 m^2_H v_d} \bigg(\sum\limits_{k= 1}^2 {V^{\star}_{k1}}\epsilon^{
u}_{k2}\tan\beta \bigg),\nonumber\\
C^H_2 &=& C^H_3 \simeq 0.
 \label{Higgsw1}
\end{eqnarray}

It should be noted that in obtaining  Eq.(\ref{Higgsw1}) several
other approximations have been taken.  First, we have dropped the
terms suppressed by the small quark masses $m_u$ and $m_d$ and the
terms suppressed by the CKM element $V_{td}$. Second, in $C^H_4$
we dropped the term proportional to $ m_c m_s$ as it is real and
thus it will not be relevant for generating weak phases. Finally,
in $C^H_1$ we dropped the terms proportional to $ m_c $ as it will
be numerically much smaller than the other terms due to the
suppression factor $V^*_{cd}/(v \sin\beta)\simeq V^*_{cd}/v \simeq
-10^{-3}$ for large $\tan\beta$ case of our interest. It should be
noted also that, for large $\tan\beta$ case $m_s /v_d= m_s /(v
\cos\beta)$ becomes large and so $C^H_4$ becomes comparable with
$C^H_1$.

 The total amplitude of
$D^0 \to K^+ \pi^-$, including Higgs contribution, can be written
as \be {\mathcal A}^{SM+H}= \bigg(C^{SM}_1+\frac{1}{N} C^{SM}_2 +
\chi^{K^+}(C^H_1-C^H_4)\bigg)X_{D^0 \pi^-}^{K^+}-
\bigg(C^{SM}_2+{1\over N} C^{SM}_1
+\frac{1}{2N}\big(C^H_1-\chi^{D^0}
C^H_4\big)\bigg)X_{K^+\pi^-}^{D^0}\label{HigsT}\ee where
  \bea \chi^{K^+}&=&{m_K^2\over (m_c-m_d)(m_u+m_s)}\nonumber\\
\chi^{D^0}&=& {m_D^2\over (m_c+m_u)(m_d-m_s)}\eea

Eq.(\ref{HigsT}) can be expressed in terms of the amplitudes $T''$
and $E''$ introduced before as: \be {\mathcal A}^{SM+H}=
V^*_{cs}V_{ud}(T''^{SM+H}+E''^{SM+H}) \label{HigsTt}\ee where \bea
T''^{SM+H} &=& \frac{G_F}{\sqrt{2}}a^{SM+H}_1
f_{K}(m^2_D-m^2_\pi)F^{D\pi}_0(m^2_{K})\nonumber\\
E''^{SM+H} &=& \frac{G_F}{\sqrt{2}}a^{SM+H}_2
f_D(m^2_{\pi}-m^2_K)F^{K\pi}_0(m^2_D)\eea

where \bea a^{SM+H}_1 &=&\bigg(a_1+\Delta  a_1+\Delta a^{H}_1
\bigg)\label{a1t}\nonumber\\
a^{SM+H}_2&=& - \bigg(a_2 +\Delta a_2+\Delta
a^{H}_2\bigg)\label{a2t}\eea

with \bea \Delta a^{H}_1&=& \chi^{K^+}(C^H_1-C^H_4)\nonumber\\
\Delta a^{H}_2 &=& \frac{1}{2N}\big(C^H_1-\chi^{D^0}
C^H_4\big)\eea

Clearly Higgs contributions affect only the short distance physics
(Wilson coefficients) leaving the strong phases unaffected.

In a recent study a lower bound on the charged Higgs mass in 2HDM
of Type II has been set after taking into account all relevant
results from direct charged and neutral Higgs boson searches at
LEP and the LHC,  as well as  the most recent constraints from
flavour physics \cite{Arbey:2017gmh}. The bound reads $
m_{H^\pm}\gtrsim 600$ GeV independent of $\tan \beta$. This bound
should be also respected in 2HDM III \cite{Crivellin:2012ye}.

For $\tan\beta = 50$ and $m_H = 600$ GeV we find that \bea \Delta
a^{H}_1&\simeq&   -0.08 \,\epsilon^{ u}_{12}  + 0.02 \, \epsilon^{
u}_{22} + 2.88 \,\epsilon^{ u}_{22}\, \epsilon^{u\,*}_{21} + 0.52
\, \epsilon^{ u}_{12}\, \epsilon^{u\,*}_{31} - 0.12\,
\epsilon^{ u}_{22} \,\epsilon^{u\,*}_{31} \nonumber\\
\Delta a^{H}_2 &\simeq&   0.20 \,\epsilon^{ u}_{12}  - 0.05\,
\epsilon^{ u}_{22}  + 0.24 \,\epsilon^{ u}_{22}\,
\epsilon^{u\,*}_{21} + 0.04 \, \epsilon^{ u}_{12}\,
\epsilon^{u\,*}_{31} - 0.01 \,\epsilon^{ u}_{22}\,
\epsilon^{u\,*}_{31} \eea

where we have neglected the terms that are proportional to
$\epsilon^{ u}_{12} \epsilon^{ u\star}_{21}$ due to the strong
constraint imposed on $\epsilon^{ u}_{12} \epsilon^{ u\star}_{21}$
from $D-\bar D$ mixing \cite{Crivellin:2013wna}. So we are left
only with $\epsilon^u_{12},\epsilon^u_{22}$ and
$\epsilon^{u\star}_{21},\epsilon^{u\star}_{31}$. The electric
dipole moment of the neutron and the observable $B\to \tau \nu$
can be used to set constraints on the coupling $\epsilon^{u}_{31}$
\cite{Crivellin:2013wna}. These constraints indicate that the
terms proportional to $\,\epsilon^{ u}_{22}\,
\epsilon^{u\,*}_{31}$ will be much smaller compared to the terms
proportional to $\,\epsilon^{ u}_{22}$ only and so these terms can
be safely neglected. For similar reason we can drop the terms
proportional to $\,\epsilon^{ u}_{12}\, \epsilon^{u\,*}_{31}$ in
comparison to the terms proportional to $\,\epsilon^{ u}_{12}$
only. Thus we get

\bea \Delta a^{H}_1&\simeq&   -0.08 \,\epsilon^{ u}_{12}  + 0.02
\, \epsilon^{ u}_{22} + 2.88 \,\epsilon^{ u}_{22}\,
\epsilon^{u\,*}_{21}  \nonumber\\
\Delta a^{H}_2 &\simeq&   0.20 \,\epsilon^{ u}_{12}  - 0.05\,
\epsilon^{ u}_{22}  + 0.24 \,\epsilon^{ u}_{22}\,
\epsilon^{u\,*}_{21}  \eea

The couplings $\,\epsilon^{ u}_{12}$ and $\,\epsilon^{ u}_{21}$
can be constrained using the process $\bar D^0 \to \mu^+ \mu^-$
\cite{Crivellin:2013wna}. The resulting bounds can be expressed in
terms of $m_{H^\pm}$ and $\tan\beta$ as \cite{Crivellin:2013wna}

\be |\epsilon^{ u}_{12,21}| \leq  3.0\times
10^{-2}\frac{(m_{H^\pm}/500 GeV)^2}{\tan\beta/50}\ee

for $m_{H^\pm} = 600$ GeV and $\tan\beta=50$ we get $|\epsilon^{
u}_{12,21}| \leq  4.32\times 10^{-2}$. These bounds indicate that
maximum values of the real and imaginary parts of $\epsilon^{
u}_{12,21}$ will be roughly of order $10^{-2}$. We proceed now to
discuss the constraints imposed on the coupling $\,\epsilon^{
u}_{22}$. The processes $D_{(s)} \to \tau \nu $, $D_{(s)} \to \mu
\nu $ can constraint the real part of $\,\epsilon^{ u}_{22}$ while
the constraints on the imaginary part of $\,\epsilon^{ u}_{22}$
are weak \cite{Crivellin:2013wna}. For $m_{H^\pm} = 600$ GeV,
$\tan\beta=50$ and assuming real $\,\epsilon^{ u}_{22}$ the
strongest bound  $ -0.3 \lesssim \epsilon^{ u}_{22} \lesssim 0.3$
has been obtained by combining the constraints from $D \to \mu \nu
$ and $D_{s} \to \mu \nu $ \cite{Crivellin:2013wna}. Regarding the
imaginary part of $\,\epsilon^{ u}_{22}$, and for $m_{H^\pm} =
600$ GeV, $\tan\beta=50$,  the constraints from the electric
dipole moment of the neutron reads $ -0.16 \lesssim \,
Im(\epsilon^{ u}_{22})\, \lesssim 0.16$ \cite{Crivellin:2013wna}.
Other processes such as $D-\bar D$ mixing and $K-\bar K$ mixing
can be used to set bounds on $\,\epsilon^{ u}_{22}$. However these
bounds are weaker than the bounds obtained from  $D_{(s)} \to \tau
\nu $, $D_{(s)} \to \mu \nu $ and the electric dipole moment of
the neutron \cite{Delepine:2012xw,Crivellin:2013wna}.

 The real parts of $\Delta a^{H}_1$ and $\Delta a^{H}_2$ are
expected to be much smaller than the SM contributions, $a_1$ and
$a_2$, and hence we can be safely neglect them and keep only the
imaginary parts required for generating the weak phases. Thus we
get

\bea \Delta a^{H}_1&\simeq& \bigg(0.02 \, Im(\epsilon^{ u}_{22}) +
2.88
 \, Im(\epsilon^{ u}_{22})\,Re(\epsilon^{u}_{21})- 2.88
 \, Re(\epsilon^{ u}_{22})\,Im(\epsilon^{u}_{21})\bigg)I \nonumber\\
\Delta a^{H}_2 &\simeq&  \bigg( 0.20 \, Im(\epsilon^{ u}_{12})  -
0.04\,Im( \epsilon^{ u}_{22})\bigg) I  \eea

where we kept the dominant terms only after keeping in mind the
bounds on $\epsilon^{ u}_{12}$, $\epsilon^{ u}_{21}$ and
$\epsilon^{ u}_{22}$. We consider now two scenarios. In the first
scenario we assume that $\epsilon^{ u}_{22}$ is pure real and the
other couplings $\epsilon^{ u}_{12}$ and $\epsilon^{ u}_{21}$ are
pure complex. In the second scenario we assume that $\epsilon^{
u}_{22}$ is pure complex and the other couplings $\epsilon^{
u}_{12}$ and $\epsilon^{ u}_{21}$ are pure real. In each scenario
we take the maximum value of  $\epsilon^{ u}_{ij}$ for $ij=
12,21,22$ allowed from the constraints discussed before. In the
first scenario we find that

\bea \Delta a^{H}_1&\simeq& - 0.037\, I \simeq \, 0.037\,{\rm
e}^{- 90^\circ\, i} \nonumber\\
\Delta a^{H}_2 &\simeq&  0.009 \, I \simeq \, 0.009 \,{\rm e}^{
90^\circ\, i}  \eea

while in the second scenario we find that

\bea \Delta a^{H}_1&\simeq&  0.023\, I \simeq \, 0.023\,{\rm e}^{
90^\circ\, i}  \nonumber\\
\Delta a^{H}_2 &\simeq&  - 0.006 \, I \simeq \, 0.006 \,{\rm e}^{
- 90^\circ\, i} \eea

The direct CP asymmetry of $D^0 \to K^+ \pi^-$, including Higgs
contributions, can be expressed as

\begin{eqnarray}
A_{CP} &=& {| {\mathcal A}^{SM+H}|^2-|\bar  {\mathcal A}^{SM+H}|^2
\over | {\mathcal A}^{SM+H}|^2+|\bar  {\mathcal A}^{SM+H}|^2 }
\simeq \kappa\, \sin(\phi^{H}_2-\phi^{H}_1)\label{acp}
\end{eqnarray}

where where $\kappa$ is given as before and the weak phases
$\phi^{H}_1$ and $\phi^{H}_2$ are defined through \be \phi^{H}_i =
\tan^{-1}\big(\frac{|\Delta a^H_i|\sin\Delta \phi^H_i}{a_i}\big)
\ee   where $\Delta \phi^{H}_i$ is the phase of $\Delta a^{H}_i$.
Thus for the first scenario we find that

\begin{eqnarray}
A_{CP}\simeq 0.01\, \kappa \simeq 5\times 10^{-3}
\end{eqnarray}

while for the second scenario the predicted CP asymmetry is

\begin{eqnarray}
A_{CP}\simeq -0.007\, \kappa \simeq - 3\times 10^{-3}
\end{eqnarray}

Clearly the predicted direct CP asymmetries in the two scenarios
are $6$ orders of magnitude larger than the SM predicted one.

\section{Conclusion \label{sec:conclusion}}

In this paper we have studied the direct CP asymmetry of $D^0 \to
K^+ \pi^- $ within standard model and two Higgs doublet model with
generic Yukawa structure. In the standard model the tree-level
amplitude has no source of the weak phases required for generating
direct CP asymmetry. As a result we derived the corrections to the
tree level amplitude generated from the box and di-penguin
diagrams. We found that these correction can generate weak CP
violating phases. However these phases are so tiny leading to a
direct CP asymmetry of order $10^{-9}$. With this tiny CP
asymmetry the decay mode $D^0 \to K^+ \pi^- $ can  serve as a
probe of new sources of weak CP violating  phases that can be
generated in new physics beyond standard model.

As an example of  new physics beyond standard model we considered
the two Higgs doublet model with generic Yukawa structure. Within
this model we have derived the Wilson coefficients corresponding
to the decay process $D^0 \to K^+ \pi^- $ of our interest. After
discussing the relevant constraints on the parameter space of the
model, relevant to the process $D^0 \to K^+ \pi^- $, we have shown
that charged Higgs couplings to quarks can lead to a direct CP
asymmetry of order $10^{-3}$.  This asymmetry is $6$ orders of
magnitude larger than the standard model prediction.

\section*{Acknowledgements}

The work of D.D. was partially support by CONACYT project CB-259228, Conacyt-SNI and Guanajuato University through  DAIP project.

\end{document}